\documentclass[prb,twocolumn,showpacs,preprintnumbers,amsmath,amssymb]{revtex4}
\usepackage{graphicx} 
\date {}
\begin{document}
\title{Heat release in clusterization of nanoparticles}
\author{A.\,I. Karasevskii}
\email{akaras@imp.kiev.ua}\author{V.\,V.
Lubashenko}\email{vilu@imp.kiev.ua}\affiliation{G.\,V. Kurdyumov Institute for
Metal Physics
 36 Vernadsky str., Kiev, 03142, Ukraine}
\date{\today}
\begin{abstract}
We show that the effect of size dependence of the melting temperature of nanocrystals may be used to govern
anomalies of thermodynamic properties of nanocrystals in the premelting range.  For example, if temperature of nanocrystals is near the melting point, their mechanical contact should result in a change of the temperature and  release of some heat.  We discuss fundamental possibility of construction of a ``heat pump'' to extract thermal energy from a low-temperature medium.
\end{abstract}
\maketitle
%
Reduction of crystal size up to the nanoscale regime  is always accompanied by modification of its phonon and electronic spectra, thus changing both electrophysical and thermodynamic properties of the system. Such effects may be so dramatic, that the state of the matter would change fundamentally.

Nanocrystals display strong size dependence of thermodynamic properties. For example, heat capacity
\cite{Lai1996,Zhang2000,Olson}, Debye temperature \cite{Yang2007}, and melting temperature \cite{Olson,Buffat,Efremov} vary with particle size even in the mesoscopic size range. As shown in Ref. \onlinecite{KL2008}, such behaviour of thermal properties of nanocrystals is caused by size-dependent rearrangement (quantization) of their phonon spectra. Such modification of the vibrational spectrum results in changes of statistical parameters of atoms composing the nanocrystal and, therefore, in size-dependent changes of average values of energetic contributions to the free energy  of the system.

A temperature $T_c$ of anharmonic instability of the crystal is especially sensitive to the above mentioned rearrangement of the eigenfrequency spectrum due to reduction of the crystal size \cite{KL2008}. Approaching $T_c$ is accompanied by nonlinear rise of some thermodynamic properties \cite{KL2002,KH2003b}  and dramatic reduction of the formation energy of crystal structure defects \cite{KL2005,KHL2005,KL2007}. These facts allow one to relate $T_c$ with the melting temperature $T_m$ and to consider anomalous behaviour of thermodynamic properties as premelting effects \cite{KL2005}. For nanocrystals, it is important that the instability point can be approached not only by increasing temperature, but via reduction of the crystal size, too.

To determine thermodynamic properties of nanocrystals in Refs.~\onlinecite{KL2008,KL2009}, a self-consistent  statistical method \cite{KL2002} was used. The method had been developed to calculate thermodynamic characteristics of simple solids in wide range of temperature and pressure using statistical distribution functions of atomic displacements. Parameters of these distribution functions were determined by phonon spectrum only. Size influence on parameters of such statistical distribution for a nanocrystal was taken into account by consideration of size-dependent rearrangement of the vibrational spectrum \cite{KL2008}. The nearest-neighbour interaction was approximated with the Morse potential with parameters determined by fitting the observed values of interatomic distance, bulk modulus, and sublimation energy at $T=0$ \cite{KL2002,KL2004}.
\begin{figure}[t]
\includegraphics[width=\columnwidth]{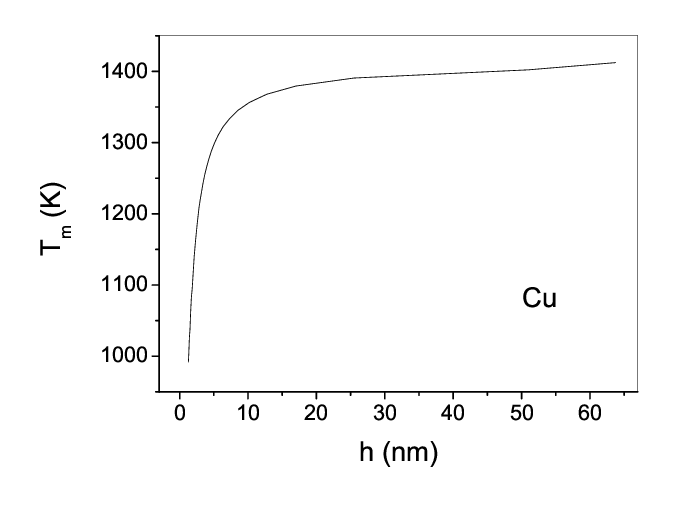}
\caption{\label{fig:1}Melting temperature of a free Cu thin plate   versus   thickness (by Ref.~\onlinecite{KL2008}).
}
\end{figure}

In Fig.~\ref{fig:1} we present the melting temperature of a Cu nanocrystalline plate versus thickness  $h$ computed in Ref.~\onlinecite{KL2008}  from the assumption that   $T_m\approx T_c$. Similar size-dependent melting curves are inherent to all kinds of free-standing nanocrystals, \textit{e.g.} rare-gas nanocrystals \cite{KL2009}, thus implying universality of the mechanism of size-dependent quantization of vibrational spectra which is responsible for a shift of temperature dependence of thermodynamical properties of nanocrystals.
\begin{figure}[ht]
\includegraphics[width=\columnwidth]{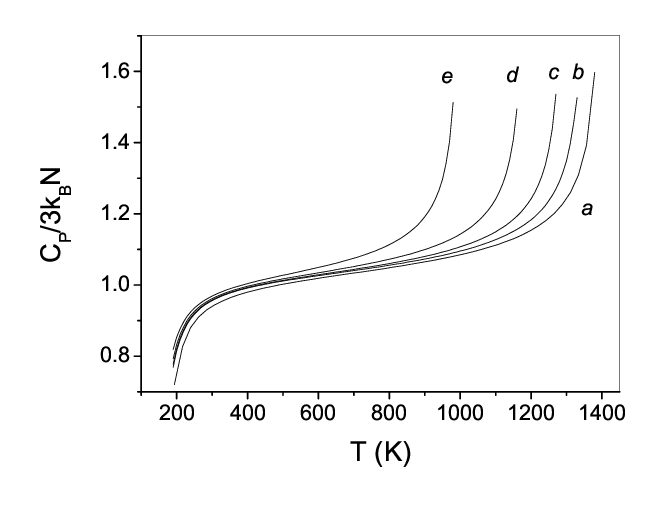}
\caption{\label{fig:2}The temperature dependence of the isobaric heat capacity of Cu nanocrystals: (a) bulk crystal; (b) $h=10.2\  \mathrm{nm}$; (c) $h=5.1\ \mathrm{nm}$;
(d) $h=2.55\ \mathrm{nm}$; (e) $h=1.26\ \mathrm{nm}$ (by Ref.~\onlinecite{KL2008}).
}
\end{figure}

Along with the melting point, premelting anomalies of thermodynamic properties  of nanocrystals also shift towards lower-temperature range. This effect is illustrated with Fig.~\ref{fig:2} which displays temperature dependence of isobaric heat capacity $C_P$  of Cu nanoplates of different sizes calculated  from the equilibrium value of the crystal free energy  \cite{KL2008}. A dramatic increase in $C_P$   corresponds to the premelting temperature range, in accord with results of calorimetric studies of nanocrystalline materials \cite{Zhang2000,Olson}. A similar behavior of $C_P(h)$ was also theoretically found for the rare-gas nanocrystals \cite{KL2009}. As crystal size decreases, the premelting anomaly takes place at lower temperatures.  Thus, size of a nanocrystal may be considered as an independent thermodynamic variable, in addition to pressure and temperature. It is important that variation of the crystal size allows one to govern the premelting temperature range, where thermodynamic properties show anomalous behaviour.

\begin{figure}[ht]
\includegraphics[width=\columnwidth]{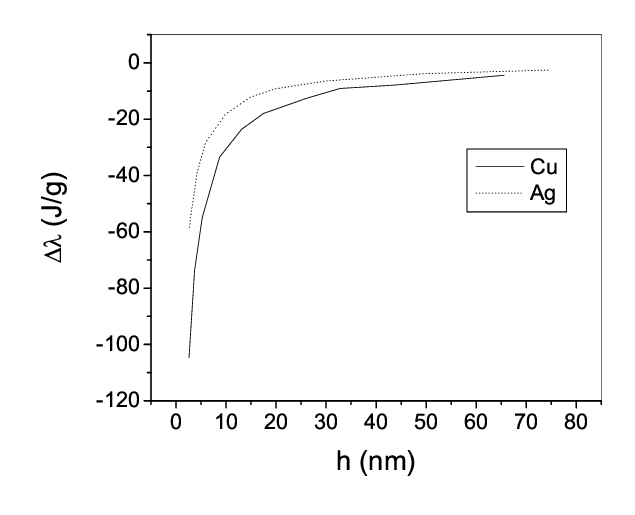}
\caption{\label{fig:3}A  change in the specific heat of fusion of a nanocrystalline plate versus thickness  calculated for Cu and Ag.
}
\end{figure}

The above size-dependent shift of the premelting  range leads to a number of physical effects that should be observed in nanocrystals. For example, a change in the specific heat of fusion $\lambda$ of a nanocrystal of size $h$ in comparison with the bulk value can be approximately represented as
\begin{equation}
\Delta \lambda=\varepsilon (T_m(h),h)-\varepsilon^0(T_m^0), \label{Deltalambda}
\end{equation}
where
\begin{equation}
\varepsilon=g-T \frac{\partial g}{\partial T} \label{epsilon}
\end{equation}
is internal energy per  mass unit, $g(T,P)$  is Gibbs free energy per   mass unit, superscript 0 denotes  bulk values of thermodynamic parameters. Using an explicit expression for $g(T,P)$ derived in Ref.~\onlinecite{KL2008}, it is easy to show that $\Delta \lambda$ depends on $h$ as
\begin{equation}
\Delta \lambda \sim T_m(h)-T_m^0 \sim \left(1+\frac{\gamma R}{h} \right)^{-3}-1, \label{Deltalambda1}
\end{equation}
where $\gamma$ is a factor taking into account discrete character of the vibrational spectrum of the nanocrystal, $R$ is interatomic distance.
In Fig.~\ref{fig:3} we plotted  $\Delta \lambda$ of a nanocrystalline plate as a function of its thickness calculated for Cu and Ag.

We may conclude from Fig.~\ref{fig:2} that an excess of specific heat accumulated in nanocrystals of smaller size should be released at tight mechanical contact of a number of particles, which is equivalent to formation of a larger particle. Such heat release should lead to a temperature change of nanocrystals and then to a heat exchange between particles and their surrounding. For quantitative description of such an effect, let us consider free nanocrystalline plates of thickness  $h \gg R$. If two such nanocrystals are brought into tight mechanical contact, the vibrational modes with wavelengths larger than a characteristic thickness of the contact become collectivized, and their frequencies become equal to the frequencies of corresponding modes of a crystal with  thickness $2h$ (see, \textit{e.~g.}, Ref.~\onlinecite{Physical_acoustics}). Breaking the contact, \textit{i.e.} separating the two thin plates, results in the reverse rearrangement of their vibrational  spectra and is accompanied by a change $\Delta \varepsilon_h$ in the specific internal energy. Obviously,  an energy change due to joining the plates is  $-\Delta \varepsilon_h$. If the nanoparticles are isolated, the energy change would lead to a change in the temperature of the nanocrystals, $\Delta T=\Delta \varepsilon_{h}/c_V$, where $c_V$ is specific isochoric heat capacity. In Fig.~{\ref{fig:4} }  we plot a contact temperature change   $\Delta T$ of two  thin plates versus reduced initial temperature  $T/T_m^0$ calculated for Cu and Ag plates of thickness $h=5.1\ \mathrm{nm}$ ($T_m(h)/T_m^0=0.919$ for Cu and 0.914 for Ag).   As follows from the results presented in Fig.~\ref{fig:4},  $\Delta T$ is maximal if the initial system of separated nanoparticles is close to the melting point. {Fig.~\ref{fig:5} represents temperature change due to contact of two non-metallic (Xe) nanocrystals  computed for two values of the plate thickness, $h=5.6\ \mathrm{nm}$ and $h=3.3\  \mathrm{nm}$.

\begin{figure}[ht]
\includegraphics[width=\columnwidth]{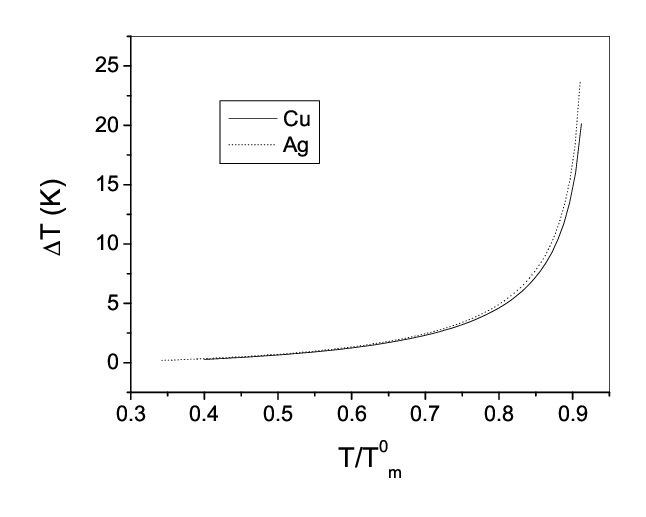}
\caption{\label{fig:4}A temperature change at mechanical contact of two  metallic thin plates of thickness $h=5.1 \ \mathrm{nm}$  versus initial temperature calculated for Cu (theoretical value of the bulk melting temperature $T_m^0=1425 \ \mathrm{K}$) and Ag ($T_m^0=1228 \ \mathrm{K}$).}
\end{figure}

\begin{figure}[ht]
\includegraphics[width=\columnwidth]{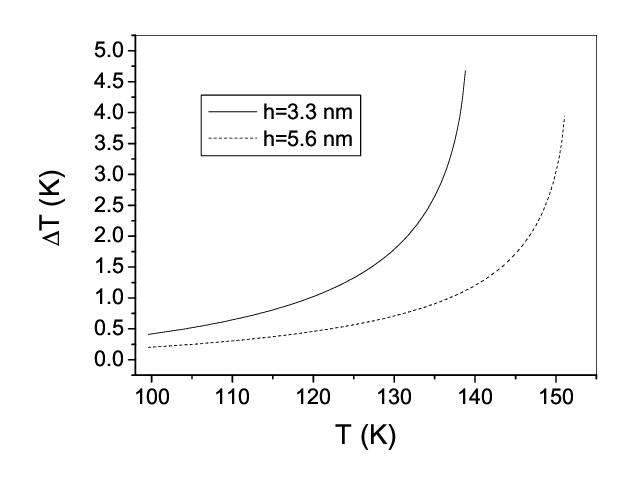}
\caption{\label{fig:5}A temperature change at mechanical contact of two Xe thin plates of thickness $h$  versus initial temperature calculated for two values of $h$.
}
\end{figure}

A contact temperature difference $\Delta T$ between a nanocrystalline cluster and surrounding medium should initiate  transport of some heat from the nanoparticles to the medium. We can estimate a transferred amount of heat from the following reasoning. Heat amount accumulated by a particle of mass $m$ and size $h$ at temperature $T$ is
\begin{equation}Q =m \int_0^T  c_P (t,h) \ \mathrm{d} t,\end{equation}
where $c_P(T,h)$ is isobaric specific heat. Usually, a change in $Q$ is ascribed to a change in temperature of the medium. However, in the case of a contact of nanocrystals whose initial temperature is close to the melting point $T_m$, there is a remarkable change of $c_P$ (Fig.~\ref{fig:2}), so that
\begin{equation}
\Delta Q \approx m\int_{0}^{T_m} [c_{P} (T,2h)-c_{P} (T,h)]\, \ \mathrm{d} T.\label{DeltaQ}
\end{equation}
Obviously, the value of  $\Delta Q$ increases substantially as the nanocrystals approach the  melting point (Figs.~\ref{fig:6}, \ref{fig:7}). For example, an estimate of the heat released in the contact of Cu plates of thickness $h=5.1 \ \mathrm{nm}$ near $T_m(h)=1310 \ \mathrm{K}$  gives about $12 \ \mathrm{J/g}$.

\begin{figure}[ht]
\includegraphics[width=\columnwidth]{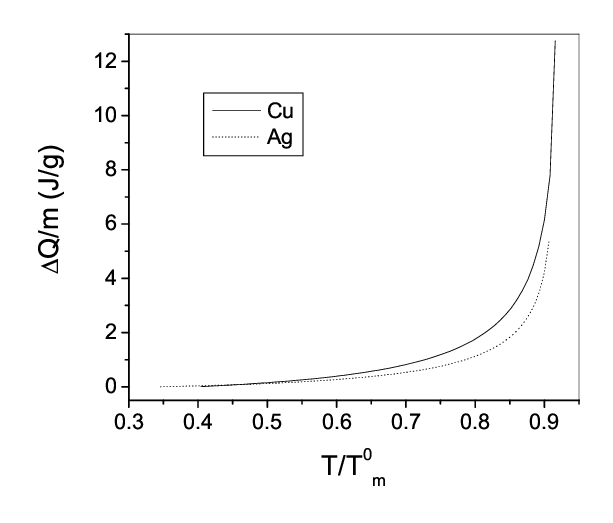}
\caption{\label{fig:6}Amount of transferred heat  versus initial temperature calculated for Cu and Ag  nanocrystals of thickness $h=5.1 \ \mathrm{nm}$.
}
\end{figure}

If  $\Delta T \sim 10-100$~K, the effect of contact change of  temperature
can be used for mechanical heat transport from media with a
lower temperature to media with a higher temperature. Indeed,
transfer of nanocrystals of size $h$ from medium 1 with
temperature $T_1$ into medium 2 with temperature $T_2>T_1$ and
their clusterization in the medium 2 results in heating of the
nanocrystals and further heat transport from the nanocrystals to
the medium 2. On the next step, returning the clustered
nanocrystals to the medium 1 and their  separation leads
to cooling of separated nanocrystals and, therefore, to heat
transport from the medium 1 to the nanocrystals. Realization of such a cyclic process would allow one to construct a ``heat pump'' to extract thermal energy from lower temperature media. It is important that, due to mechanical work performed over the system, this scheme does not contradict to the  second law of thermodynamics: ``Heat generally cannot flow \textit{spontaneously} from a material at lower temperature to a material at higher temperature'' (Clausius statement).

\begin{figure}[ht]
\includegraphics[width=\columnwidth]{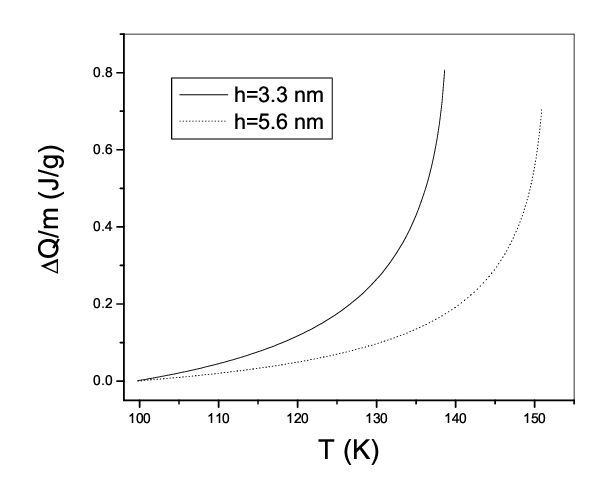}
\caption{\label{fig:7}Amount of transferred heat  versus initial temperature of Xe nanocrystals calculated for two values of $h$.
}
\end{figure}

The effect of heat release and change of temperature due to contact of nanoparticles may play important role in biological processes, \textit{e.g.} as a channel for viruses to obtain energy from the environment.

To conclude, it is important to emphasize that the present study is focused on consideration of thermal effects in nanocrystals of mesoscopic size range, so we can neglect effects due to structural changes inherent in nanoclusters. Such structural transitions are observed, \textit{e.g.}, in Au nanoclusters \cite{Cleveland,Nam,Koga} and result in hysteresis phenomena at melting. It should be also noted that size-dependent thermodynamic effects in nanocrystals are caused by quantization of phonon spectra and have universal nature. Contribution of conduction electrons to size effects is proportional to a factor of $\lambda_F/h$ \cite{Krivoglaz}, where $\lambda_F$ is electron wavelength at the Fermi level. For metals,  $\lambda_F/h \ll 1$, therefore, electronic influence on size-dependent changes of thermodynamic properties can be neglected. This contribution may be important, \textit{e.g.}, in nanocrystalline degenerated semiconductors \cite{Krivoglaz} where $\lambda_F/h \lesssim 1$.

\section*{Acknowledgement}
This  work was supported in part by Award no.~28/09-H  in the
framework of the Complex Program of Fundamental Investigations
``Nanosized systems, nanomaterials, nanotechnology'' of National
Academy of Sciences of Ukraine.

\end{document}